# Multi-ion conduction bands in a simple model of calcium ion channels

I. Kaufman[1], D.G. Luchinsky[1,2], R. Tindjong[1], P.V.E. McClintock[1*] and R.S. Eisenberg[3]

[1] Department of Physics, Lancaster University, Lancaster, LA1 4YB, UK;
[2] NASA Ames Research Center, MS 269-3, Moffett Field, CA, 94035, USA;
[3] The Department of Molecular Biophysics and Physiology, Rush Medical College, IL 60612, USA

*E-mail: p.v.e.mcclintock@lancaster.ac.uk;

We report self-consistent Brownian dynamics simulations of a simple electrostatic model of the selectivity filters (SF) of calcium ion channels. They reveal regular structure in the conductance and selectivity as functions of the fixed negative charge $Q_f$ at the SF. This structure comprises distinct regions of high conductance (conduction bands) M0, M1, M2 separated by regions of almost zero-conductance (stop-bands). Two of these conduction bands, M1 and M2, are related to the saturated calcium occupancies of $P=1$ and $P=2$, respectively and demonstrate self-sustained conductivity. Despite the model's limitations, its M1 and M2 bands show high calcium selectivity and prominent anomalous mole fraction effects and can be identified with the L-type and RyR calcium channels. The non-selective band M0 can be identified with a non-selective cation channel, or with OmpF porin.

## 1 Introduction

Voltage-gated calcium ion channels play an important role in stimulating muscle contraction, in neurotransmitter secretion, gene regulation and transmission of action potentials, based on their high selectivity for divalent calcium ions $Ca^{2+}$ over monovalent sodium ions $Na^+$. They exhibit the anomalous mole fraction effect (AMFE), an effective blockade of $Na^+$ permeation by small concentrations of $Ca^{2+}$, combined with measurable $Ca^{2+}$ currents in the pA range [1].

The selectivity of calcium channels is defined by a narrow selectivity filter (SF) with a strong binding site formed by negatively-charged protein residues [1, 2]. Wild-type calcium channels and their mutants differ in the composition, structure (locus) and net fixed charge $Q_f$ of these protein residues at the SF. The most-studied L-type calcium channel possesses an EEEE locus with an estimated $Q_f$=3-4$e$ [2, 3], where $e$=-1.6x10$^{-19}$C is the electronic charge. The ligand-gated Ryanodine receptor (RyR) calcium channel has a DDDD locus with a larger $Q_f \approx 4.5e$ [4]. The L-type and RyR channels exhibit different threshold concentrations for blockage of $Na^+$ current by $Ca^{2+}$ ions: $[Ca]_{50} \approx 1 \mu M$ and $[Ca]_{50} \approx 1 mM$ respectively [4].

Sodium channels have structures very similar to calcium channels, but with different SF loci (and therefore different $Q_f$), and different lengths and radii [5-7]. The eukaryotic sodium channel has a DEKA locus with $Q_f \approx 1e$ [5, 7]. In the recently studied bacterial Navab channel, the four Glutamate side chains form a 6.5x6.5Å scaffold with an orifice of 4.6x4.6Å defined by van der Waals surfaces [6, 8].

Mutant studies show that $Q_f$ is crucial in determining the Ca vs Na selectivity of calcium channels. Usually, mutations that influence $Q_f$ also destroy the channel's selectivity and hence physiological functionality [9]. However, an appropriate point mutation of the DEKA sodium channel ($Q_f \approx 1e$) converts it into a calcium-selective channel with a DEEA locus and $Q_f \approx 4e$ [10]. The essentially nonselective bacterial OmpF porin ($Q_f \approx 1e$) can be also turned into a Ca-selective channel by introduction of two additional glutamates in the constriction zone. The resultant mutant contains a DEEE-locus ($Q_f \approx 4e$) and exhibits an Na current with a strongly increased sensitivity to 1mM Ca [11].

Dynamic Monte-Carlo simulations of the flexible volume exclusion model of calcium and sodium channels [3, 12] show that the charge density at the SF is the first-order determinant of selectivity [3];

and the $Na^+$ to $Ca^{2+}$ occupancy ratio decreases monotonically as $Q_f$ increases from $1e$ (DEKA locus) to $4e$ (DEEE locus), while the pore becomes more and more $Ca^{2+}$ selective [5].

Self-consistent Brownian dynamics (BD) simulations [2] of conductance and selectivity, based on a purely electrostatic model of the L-type channel with a rigid binding site, revealed a narrow peak in $Ca^{2+}$ conductance near $Q_f = 3.2e$. [2, 13]

Although mutant studies [9-11] and simulations [2, 3, 5] have demonstrated very clearly the dominant influence of $Q_f$ on the selectivity of the calcium channel, it has remained unclear why particular values of $Q_f$ should be optimal for selectivity or how many such values may exist.

Multi-ion knock-on conductivity is assumed to be one of the main mechanisms of permeation and selectivity in the case of the potassium channel [14-17]; a similar mechanism could resolve the paradox of high selectivity and high conductance in the calcium ion channel [18, 19]. Its selectivity and conductivity are connected with strong ion-ion repulsion and multi-ion occupancy of the SF [2, 12, 19]. The discreteness of the ionic occupancy also plays a significant role in conduction and is expected to manifest as discrete multi-ion steps/bands in the dependences of occupancy and conductance on concentration, time and other parameters [20-22]. These steps/bands are related to the barrier-less knock-on conduction mechanism that has been suggested as being the underlying mechanism responsible for the conductivity and selectivity of ion channels [17]. Simplified ("toy") electrostatic models of an ion channel, describing it as a water-filled charged protein hub in the cell membrane, were found to reproduce significant features related to the conductivity and, in particular, to the valence selectivity between ions of different charge [13, 22-25].

In this article, we show that the $Ca^{2+}$ conduction and $Ca^{2+}/Na^+$ valence selectivity of a simple calcium channel model forms a regular structure of conduction/selectivity bands (regions) as a function of $Q_f$, separated by non-conduction bands, related to saturated, barrier-less, conductivity with different numbers of $Ca^{2+}$ ions involved in the conduction. The conductance peak obtained in [2] is one part of this structure. We infer that all calcium-selective channels (both wild-type and mutants) should correspond to one of these bands.

## 2 Methods

### 2.1 A simple model of the calcium channel

We use a simple, self-consistent, purely electrostatic model of a calcium ion channel to study the effects of surface charge $Q_f$ on its conduction and selectivity. This model represents the channel's selectivity filter (SF) as a negatively-charged, axisymmetrical, water-filled protein hub in the cell membrane (Figure 1) similar to that used in earlier work [2, 3, 26]. We model the SF only following [3, 26]: we omit the charged vestibule modelled in [2] because the ion-ion and ion-fixed-charge interactions inside the SF are the main determinants of selectivity [2, 3]. The negatively-charged protein residues are modeled as a single, thin, uniformly-charged, centrally-placed, rigid ring around the SF. Located inside the wall, the ring brings a net negative charge $Q_f$ of 0- $6.5e$. The extracellular (left) and intracellular (right) baths are filled with ionic sodium-only, calcium-only, or mixed sodium-calcium aqueous solutions. In what follows we assume an asymmetrical ionic concentration: $C_L > 0$, $C_R = 0$.

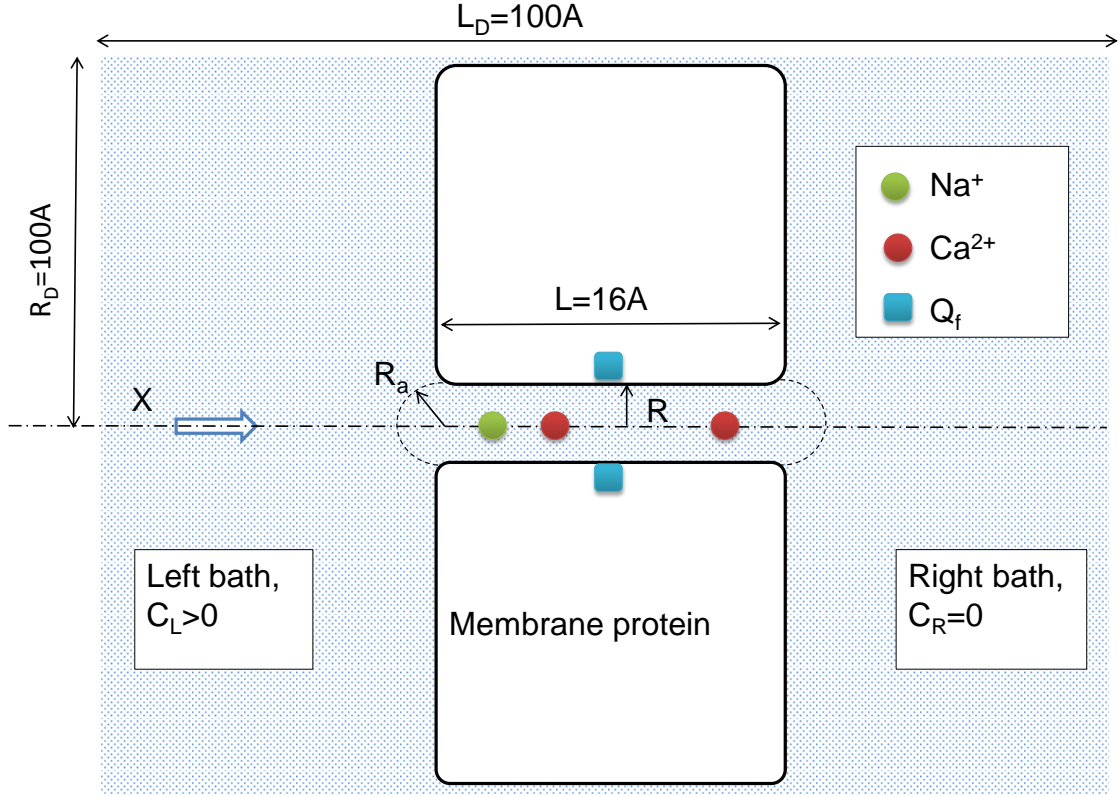

Figure 1. Computational domain for a simple model of the calcium ion channel. Its selectivity filter is treated as an axis-symmetrical water-filled cylindrical hole of radius $R$=3Å and length $L$=12-16Å through the protein hub in the cellular membrane. There is a centrally-placed, uniformly-charged, rigid ring of negative charge $Q_f$=0-6.5$e$. The left-hand bath, modeling the extracellular space, contains non-zero concentrations of $Ca^{2+}$ or $Na^+$ ions. These are injected at the Smoluchowski diffusion rate at radius $R_a$ (dashed hemispheres), we take $R_a$=$R$ here. The domain length is $L_D$=100Å, the domain radius is $R_D$=100Å, the grid size is $h$=0.5Å, and a potential difference of 0-75mV is applied between the left and right domain boundaries.

The minimum possible radius $R$ of the SF of an L-type calcium channel has been determined as being $R$=2.8Å. We use the value of $R$=3.0Å. We choose the length $L$=16Å for the main simulations as being close to the value used in [3]. Moving sodium and calcium ions are described as charged particles of radius $R_i$≈1Å (matching both ions), with diffusion coefficients of $D_{Na}$=1.17x10⁻⁹ m²/s and $D_{Ca}$=0.79 x 10⁻⁹ m²/s, respectively. The domain length is $L_d$=100Å, the domain radius is $L_r$=100Å, and the grid size is $h$=0.5Å. A voltage $V$=0-75mV was applied between the left and right domain boundaries.

### 2.2    Self-consistent electrostatics of the ion channel

The electrostatic field was derived by numerical solution of Poisson's equation:

$$-\nabla(\varepsilon\nabla u) = \frac{\sum ez_i n_i}{\varepsilon_0} \qquad (1)$$

where $\varepsilon_0$ is the dielectric permittivity of vacuum, $\varepsilon$ is the dielectric permittivity of the medium (water or protein), $u$ is the electric potential, $e$ is the elementary charge, $z_i$ is the charge number, and $n_i$ is the number density of ions.

Self-consistent electrostatic potentials and forces are calculated numerically with respect to the ion-ion interaction, channel geometry (Figure 1), self-potential barrier of the dielectric boundary force [23], value of $Q_f$ and value of applied potential $V$ at each simulation step. We used a finite volume Poisson solver specially designed for use within the ion channel geometry where there are severe jumps in dielectric permittivity. The linearity of Poisson's equation together with the superposition principle allow us to use pre-calculated lookup tables for the field components, thereby dramatically decreasing the computation time of BD simulation steps [27].

Our reduced model of the SF takes both water and protein to be homogenous continua with dielectric constants $\varepsilon_w$=80 and $\varepsilon_p$=2, respectively, together with a primitive model of ion hydration (the validity of this latter approximation is discussed below). The self-consistent electrostatics of a narrow, water-filled, channel in the protein wall differs significantly from bulk electrostatics, even when the dielectric constant of the water inside the channel is taken to be the same as in the bulk [28]. The huge gradient between $\varepsilon_w$=80 and $\varepsilon_p$=2 results in the quasi-1D axial behaviour of the electrostatic field, and hence in single-file movement of positive ions inside the channel [25]. This effect can be interpreted as electrostatic amplification of the electric field inside the channel [25, 29]. The electrostatics of the ion channel also prohibits the entrance of any negatively charged ions due to combined influences of the dielectric boundary force and the interaction with the fixed charge [23]. Consequently, we use a 1D dynamic model to simulate the axial, single-file, movement of cations (only) inside the SF and in its close vicinity.

This generic model of the ion channel can be used to describe the SF of both calcium and sodium channel depending of selected values of $R$, $L$ and $Q_f$ [3, 5]. A similar model of the Gramicidin A channel is described in [30, 31].

## 2.3 Brownian dynamics simulations

The BD simulations are based on numerical solution of the 1D over-damped Langevin equation:

$$x' = -Du_x + \sqrt{2D}\,\xi(t) \tag{2}$$

where $x$ stands for the ion's position, $D$ is its diffusion coefficient, $u$ is the self-consistent potential in $k_BT/e$ units, and $\xi(t)$ is normalized white noise. Numerical solution of (2) has been implemented with the Euler forward scheme.

We use an ion injection scheme that allows us to avoid heavy-duty simulation of ionic movements in the bulk liquid. Ions are injected randomly into the vicinity of the left channel entrance at an arrival rate that simulates the diffusive ionic flux from the undisturbed bulk. The model includes a hemisphere of radius $R_a$ at each entrance representing the boundary between the channel vicinity and the baths (we take $R_a$=$R$ here). The arrival rate $j_{arr}$ is connected to the bulk concentration $C$ through the Smoluchowski diffusion rate: $j_{arr} = 2\pi D R_a C$ [32].

The motion of each injected ion is simulated in accordance with (2) until it reaches a domain boundary, where it is assumed to be absorbed. Simulations continue until the chosen simulation time has been reached. The ionic current $J$ is calculated as averaged difference between the numbers of similar ions passing the central cross-section of the SF per second in the forward and reverse direction [17].

A number of quantities is measured during the simulations, including the sodium $J_{Na}$ and calcium $J_{Ca}$ ion currents, the partial ionic occupancy profiles $P_s(x)$ along $x$ for different concentrations, and the partial $P_{Na}$ and $P_{Ca}$ occupancies, in each case as functions of the respective concentrations of calcium *[Ca]* or sodium *[Na]*.

The BD simulations of ion current $J$ and occupancy $P$ were performed separately for $CaCl_2$ and NaCl solutions, and also for a mixed-salt configuration, with concentrations *[Na]*=30mM and 20µM ≤ *[Ca]* ≤ 80mM. The value of $Q_f$ was varied within the range 0-6.5e in order to cover the known variants of sodium and calcium channels [5].

### 2.4 Model validity and limitations

The validity of continuum electrostatics and Langevin dynamics for a narrow water-filled pore is, of course, a highly significant question in relation to simplified models of the kind discussed [33-35]. The effective screening inside the long, narrow, selectivity filter could be very different than that assumed for $\varepsilon_w$ = 80 and the whole notion of a dielectric constant inside a cylinder with $R$=3.0 Å with $L$=16 Å could become ill-defined [34].

The validity of both the electrostatics and the dynamics depends on the degree of dehydration of the ion inside the channel, so it can be defined roughly by the relationship between the channel radius $R$ and the radius of the ion's first hydration shell $R_h$ [28]. Continuum electrostatics and dynamics generally fail when $R_h > R$, but still can be applied for $R_h \approx R$ provided that one uses *effective values* of $\varepsilon_w$ and the diffusion coefficients $D_{Na}$, $D_{Ca}$ that are all dependent on $R$ [28]. We estimate $R_h \approx 3.5$Å for $Na^+$ and $Ca^{2+}$ ions so the calcium channel with $R \approx 3$ Å [1] does provide some room for $Na^+$ and $Ca^{2+}$ ions to carry water molecules. Both ions are still partially hydrated, therefore, and the continuum approximation with effective values can be used inside the SF. It is shown in [28] that the effective $\varepsilon_w$ saturates to its bulk value $\varepsilon_w$=80 for $R \approx 3.5$ Å (roughly corresponding to $R_h$) and is still close to it ($\varepsilon_w \approx 70$) for $R$=3Å. This allows us to use bulk value for $\varepsilon_w$. The effective values of the ionic diffusion coefficients also decreased significantly with decreasing $R$ in comparison with their bulk values, and are estimated as $D \sim 0.25 D_{bulk}$ for $R$=3Å [34].

In this research we therefore use the bulk values of $\varepsilon_w$ and $D$ as their effective values throughout the whole computational domain, including the SF, a choice that avoids the use of additional fitting parameters. This reduced model obviously represents a considerable simplification of the actual electrostatics and dynamics of moving ions and water molecules in single-file within the narrow SF [34, 36]. Nonetheless, its applicability is supported by its reproduction of the experimentally observed AMFE kinetics for the calcium channel (see below, Figures 4 and 5).

We have performed a parametric study of the stability and variability of the simulation results to changes of the effective model parameters: the length $L$ and radius $R$ of the SF, the length $H$ of the charged ring, and the applied voltage $V$ (see Appendix). We show that the simulation results are relatively insensitive to changes in the model parameters and that the main result – the distinct conductivity and selectivity bands – exists within a range of parameter values. This stability implicitly confirms that reported phenomena arise from the basic electrostatics of real ion channels rather than as computational artifacts of our model related to some particular values of parameters.

We note that the recent determination of the crystal structure of the sodium NavAb channel [6] provides for well-hydrated motion of ions and has a rather asymmetric mode of coordination by four residues [8]. Although this structure refers to a bacterial channel, and not a DEKA-motif channel, it may nonetheless be relevant to the Na- and Ca-selective sites in the channels under consideration. If this turns out to be the case, the model presented here, which assumes a symmetrical arrangement of fixed charges and ions inside the channel, can with advantage be modified accordingly.

## 3 Results and Discussion

### 3.1 Conduction and selectivity bands of the calcium channel

Figure 2 (a) and (b) demonstrate our main result - the appearance of regular structure in the $Ca^{2+}$ ion current $J$ as a function of $Q_f$, and *[Ca]*, comprising areas of high conductance (conduction bands) M0, M1, M2 separated by almost zero-conductance stop-bands. The peak separation $\Delta Q \approx 2e$ corresponds roughly to the charge on one $Ca^{2+}$ ion. Band M1 coincides with the $J$ peak from [2] which was obtained for a negative charge of $1.3 \times 10^{-19}$ C/ glutamate residue giving a total charge of $Q_f = 4 \times 1.3 \times 10^{-19}$C$= \sim 3.2e$ (the authors of [2] used 4 residues to model the EEEE locus of an L-type calcium channel).

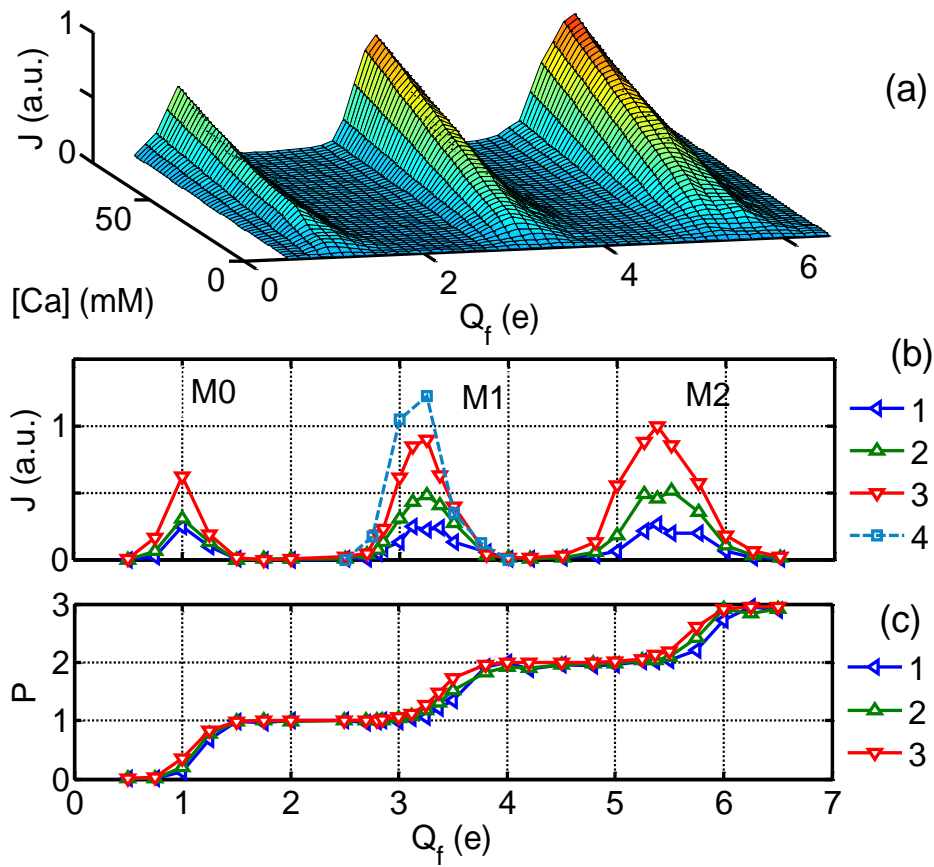

Figure 2. Multi-ion conduction bands of the calcium channel model. (a) A 3D plot of the calcium current $J$ vs fixed charge $Q_f$ and concentration *[Ca]* exhibits regular band structure. (b) A plot of $J$ as a function of $Q_f$ and *[Ca]* shows the M0, M1, and M2 bands: plots 1,2,3 are $J$ and $P$ for *[Ca]*=20mM, 40mM and 80mM, respectively; and 4 is the $J$ peak from [2], corresponding to M1. (c) The occupancy $P$ shows stepwise growth as $Q_f$ increases. The flat steps correspond to the saturated occupancy values $P$=1, 2, 3, ..

The $Ca^{2+}$ occupancy $P$ exhibits step-wise growth with increasing $Q_f$ (Figure 2 (c)). The flat steps correspond to non-conducting saturated states with $P$=0, 1, 2... where the potential well at the binding site is too deep to allow escape of a $Ca^{2+}$ ion, and too shallow to allow the next $Ca^{2+}$ ion to enter the SF and push the bound ion(s) out. The conduction bands M0, M1 and M2 correspond to transitions in $P$:

$0\rightarrow1$, $1\rightarrow2$, and $2\rightarrow3$, respectively. This picture corresponds to the "knock-on" mechanism of $Ca^{2+}$ conductance and selectivity [1, 2]. It has an obvious analogue in semiconductor physics, where conduction also occurs in partially filled bands [37].

The appearance of the distinct conduction bands is caused by the discreteness of the multi-ion occupancy $P$. Their existence just for $Ca^{2+}$ in the calcium channel relates to the high $Q_f$ and double-valence of $Ca^{2+}$, enhancing the electrostatic effects of valence selectivity [13].

The effective parameter values in our model ($R$, $L$) can be varied within restricted ranges to fit experimental data for real channels [2, 3]. Our parametric study (see Appendix) shows that the simulated bands are only weakly sensitive to variations of $R$ and $L$ in the ranges: $R$=2.5-3.5Å, $L$=12-20 Å. Further decrease of $L$, or increase of $R$, leads to a weakening of the bands and to their eventual disappearance (at $L$=8Å for $R$=3Å). Their disappearance for shorter SFs may be related to the existence of a minimum $L$ needed to hold two $Ca^{2+}$ ions in the SF against their mutual Coulomb repulsion.

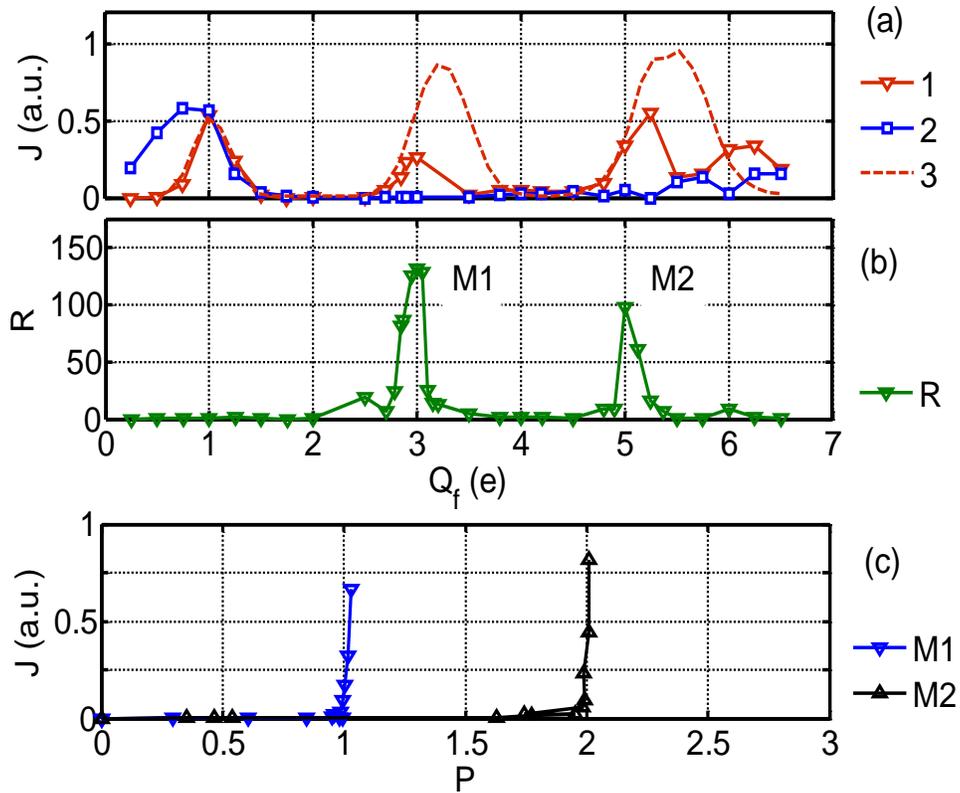

Figure 3. Conduction and selectivity bands for a *[Na]*=30mM, *[Ca]*=40mM mixed salt bath. (a) Currents *J vs* fixed charge $Q_f$. Curves: 1 - Ca, 2 - Na, 3 – Ca for a pure bath (reference curve from Figure 1). (b) The selectivity ratio $R_S$=$J_{Ca}/J_{Na}$ exhibits sharp peaks for the M1 and M2 bands. (c) $J_{Ca}$ *vs* $Ca^{2+}$ occupancy $P$. The selectivity peaks M1, M2 show saturated conductance at nearly constant $P$.

Mixed salt simulations (Figure 3 (a)) show that the M1 and M2 $J_{Ca}$ peaks are decreased and shifted to the beginning of the transition regions in $P$, as compared to the corresponding peaks for a pure $Ca^{2+}$ bath, due to attenuation by $Na^+$ ions [2]. Figure 3 (b) shows that the selectivity ratio $R_s$=($J_{Ca}/J_{Na}$) peaks at M1 and M2, with $R_s\approx130$ for the M1 conduction band, and that there is no selectivity outside these bands. The $J$ vs $P$ plots (Figure 3 (c)) are each drawn as a $J$ *vs* *[Ca]* simulation at constant $Q_f$. They show that, at $P$=1 for the M1 mode, and $P$=2 for M2, $P$ remains almost constant (saturated) while $J$ increases very rapidly. The saturated conductance appears only for distinct values of $Q_f$ where the output barrier for the bound $Ca^{2+}$ ion falls close to zero due to Coulomb repulsion between the $Ca^{2+}$ ions, leading to barrier-less conduction [17, 22]. The arriving ion instantly pushes out and replaces the bound ion, so that $P$ remains constant [17]. The saturated conductance provides the highly selective self-sustained $Ca^{2+}$ flux in a mixed salt bath with a permanently $Ca^{2+}$-occupied channel that is blocked for $Na^+$ ions [2].

### 3.2 AMFE kinetics and identification of conduction bands

Because the ion channels in living cells are designed to conduct particular ions selectively with high rates, we expect wild-type channels to correspond to one of the highly conductive or highly selective bands: M0 (non-selective), M1 or M2 (calcium selective). We try to identify these bands with known wild-types channels by comparing their simulated and measured conductivity and selectivity properties (see Table).

The conductivity band M0 at $Q_f \approx 1e$ provides non-selective conductance for both $Ca^{2+}$ and $Na^+$ ions and does not exhibit mutual blockade in AMFE simulations. Such properties could correspond to a non-selective cation channel [38] or the OmpF channel with $Q_f \approx 1e$ [11].

The highly calcium selective bands M1 and M2 correspond to different modifications of the calcium channel. We now argue that, based on their $Q_f$ values and AMFE properties, the M1 and M2 bands correspond respectively to the L-type and RyR calcium channels.

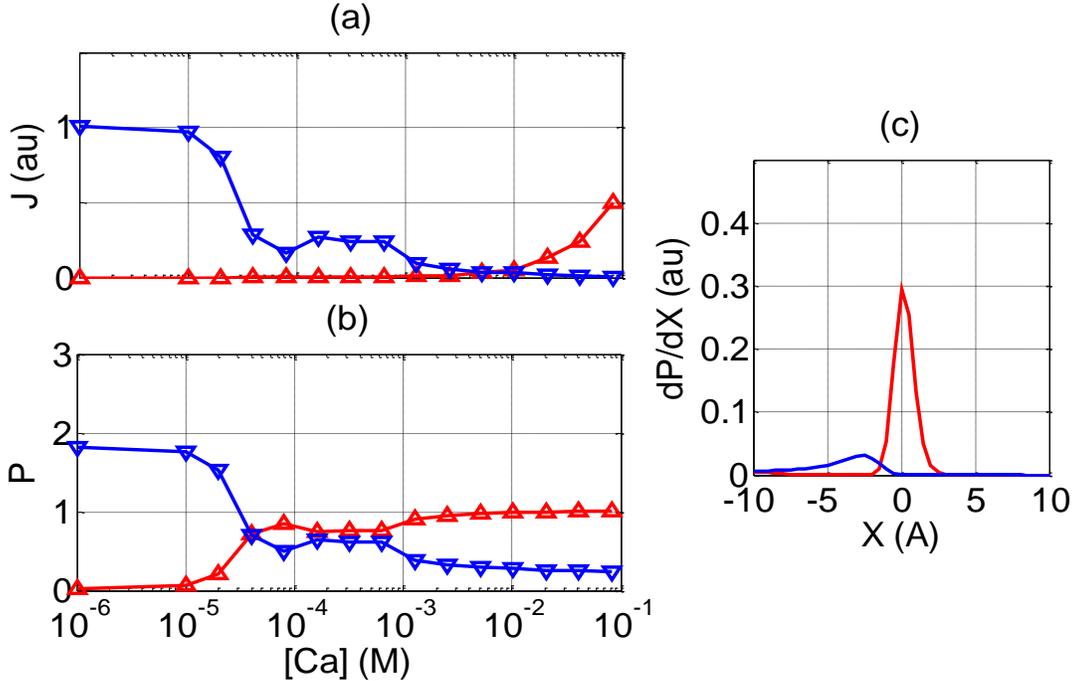

Figure 4. AMFE in a mixed salt bath for the M1 channel. (a),(b) Sodium (blue, point-down, triangles) and calcium (red, point-up, triangles) currents $J$ and occupancies $P$ $vs$ $Ca^{2+}$ concentration *[Ca]* in the highly-selective M1 channel for *[Na]*=30mM. The lines are guides to the eye. M1 shows strong blockade and AMFE at $P_{Ca}$=1 with a threshold of *[Ca]$_{50}$*≈ 30μM. (c) Mutual occupancy profiles for Na (left, blue, curve) and Ca (right, red, curve) ions show blockade of Na ions by the first Ca ion.

Figure 4 presents the dependences of $J$ and $P$ on $[Ca]$ for the M1 band in a mixed salt configuration. As shown in (a),(b), the M1 band with $Q_f=3e$ shows a strong blockade of the current $J_{Na}$ of Na$^+$ ions with a blockade onset at $[Ca]_{50} \approx$ 30μM. The blockade occurs after the first Ca$^{2+}$ ion has occupied the SF (Figure 4 (c)): $P_{Ca} \to 1$. The strong blockade with relatively low onset agrees qualitatively with the observed properties of the L-type channel [1]. The value of $Q_f$ and the conduction mechanism for M1 also correspond to the model [2] of the L-type channel (EEEE locus).

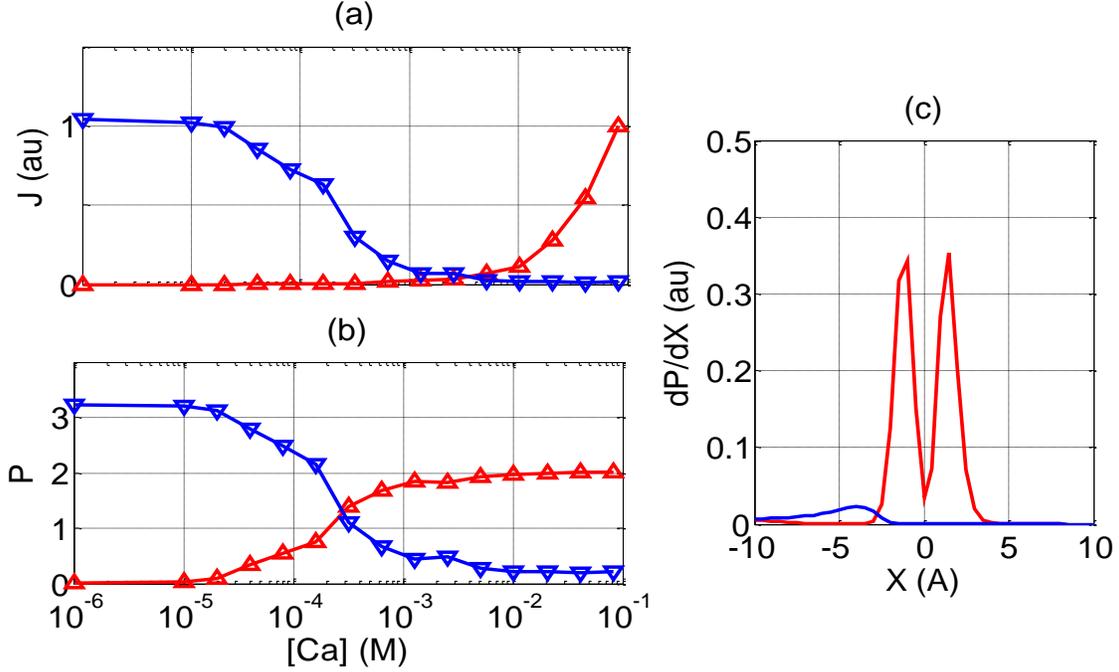

Figure 5. AMFE for a mixed salt bath for M2 channel. (a),(b) Sodium (blue, point-down, triangles) and calcium (red, point-up, triangles) currents $J$ and occupancies $P$ *vs* Ca$^{2+}$ concentration $[Ca]$ in-selective M2 channels for $[Na]$=30mM. The lines are guides to the eye. M2 shows strong blockade and AMFE at $P_{Ca}=2$ with threshold $[Ca]_{50} \approx$ 200μM. (c) Mutual occupancy profiles for Na (left, blue, curve) and Ca (right, red, curve) ions show blockade of Na ions by two Ca ions.

Figure 5 presents AMFE kinetics for the M2 band with $Q_f=5e$. Blockade onset occurs at $[Ca]_{50} \approx$ 0.2mM, corresponding to double-occupancy of the SF. Figure 5 (c) shows that the Na$^+$ ion is blocked by two Ca$^{2+}$ ions bound to the SF and separated by ion-ion repulsion. The M2 channel also shows a larger calcium current than M1. The high blockade offset and high calcium current, together with the higher value of $Q_f$, can be matched to the RyR channel (DDDD locus) [4].

The identification results are collected in Table 1.

Table 1. Identification of conduction and selectivity bands

| Conduction band | Fixed protein charge | Na conductivity | Ca conductivity | Blockade of Na current by Ca ions | AMFE | Channel(s) | Residues locus |
|---|---|---|---|---|---|---|---|
| M0 | $\approx 1e$ | High | High | No blockade | No AMFE | Non-selective cation channel [38], OmpF channel [11] | |
| M1 | $\approx 3e$ | High | High | Yes, sodium ion is blocked by one calcium ion, low blockade offset (30μM) | Yes, blockade is followed by moderate calcium current | L-type Calcium channel [1] | EEEE |
| M2 | $\approx 5e$ | High | High | Yes, sodium ion is blocked by two calcium ions, higher blockade offset (200μM) | Yes, blockade is followed by strong calcium current | RyR Channel [4] | DDDD |

## 4    Conclusions

In conclusion, our self-consistent BD simulations in a reduced model of a calcium channel SF have revealed distinct conduction bands M0, M1 and M2 as a function of the fixed charge $Q_f$ related respectively to integer values 0,1,2 of the occupancy $P$. The M0 band appears at $Q_f=1e$ with $P=1$; the M1 band appears at $Q_f=3e$ with $P=1$; and the M2 band appears at $Q_f=5e$ with $P=2$. The M0 band exhibits non-selective conductivity for both calcium and sodium ions, and can be identified with OmpF channel. The M1 and M2 bands show saturated self-sustained conductivity with high calcium selectivity ($R_s=J_{Ca}/J_{Na}\approx130$ for the M1 band) and prominent AMFE, and can be identified with the L-type (EEEE locus) and RyR (DDDD locus) calcium channels, respectively. The existence of the bands also provides a possible explanation for the results of mutation studies [9-11] in which a change in the fixed charge was found to destroy functionality or alter the type of selectivity.

Finally, we speculate that gating, and drug-induced blockade, may correspond to switching between a conduction band and a stop-band [20].

## 5    Acknowledgements

This work was supported by Engineering and Physical Sciences Research Council (EPSRC) (grant No: EP/G070660/1).

## 6    Appendix: Parametric study

We have performed a parametric study of the stability and variability of the simulated band structure (see above) in the face of changes in the effective model parameters: the length $L$ and radius $R$ of the SF, the length $H$ of the charged ring, and the applied voltage $V$ (see Figure A1 -Figure A4 below). In fact, the characteristic structure exists within a range of parameter values, as we now describe in more detail.

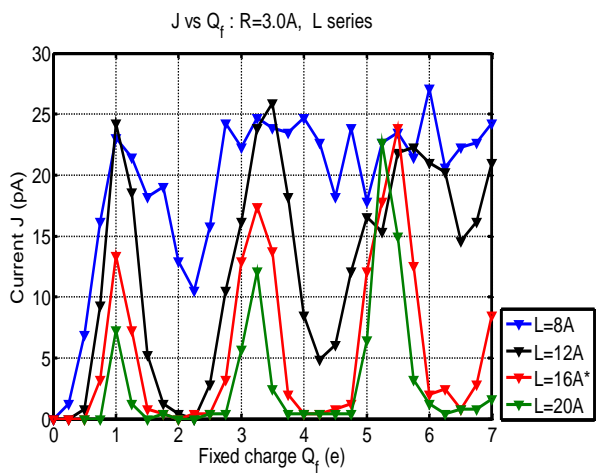

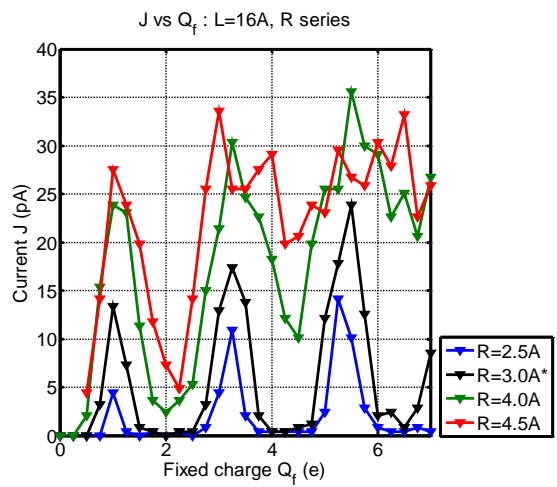

Figure A1. Calcium current $J$ vs fixed charge $Q_f$, showing how the band structure changes as the length of the SF varies in the range $L$=8-20 Å. Increase of $L$ leads to an increase of the contrast between conduction/non-conduction bands, combined with a decrease of $J$, and vice versa.

Figure A2. Calcium current $J$ vs fixed charge $Q_f$ showing how the band structure changes with the radius of the SF in the range $R$=2.5-4.5Å. An increase of $R$ to 3-4Å leads to a decrease in the bands' contrast, but the general pattern of the bands is still visible. A further increase of $R$ to 4.5Å destroys the band structure.

The value of length $L$ for the SF of calcium channels is usually estimated and modelled as being in the range $L$=5-15Å [2, 3] but the *effective* length could also depend on the geometry of the SF vestibules. We have performed BD simulations with $L$ varying from 8-20 Å to investigate the parametric dependence of the conductance bands on $L$, as shown in Figure A1. An increase in $L$ leads to a sharpening of the band structure and to decreasing conductance. A decrease in $L$ leads to the flattening and eventual disappearance of the band structure (at $L$<=10Å for $R$=3Å) combined with an increase in conductance.

The effect of varying $R$, shown in Figure A2 reveals a significant decrease of selectivity with increasing $R$, and vice-versa. The band structure eventually disappears at $R$=4.5Å.

Figure A3 shows the result of varying the length $H$ of the charged ring within the range 0-8Å. It is evident that the band structure is relatively insensitive to $H$. This finding agrees with the results of dynamic Monte-Carlo simulations [3, 5] for a flexible SF model, which showed that the selectivity is defined by the net charge rather than by the axial distribution of fixed charges.

The results in Figure A4 show that the band structure is relatively insensitive to variations in the applied voltage $V$ within the range 0-75mV.

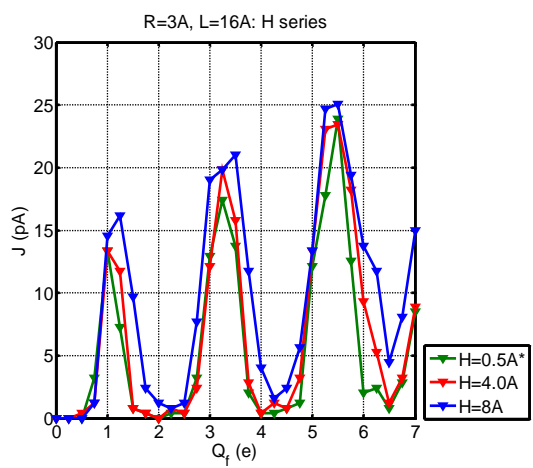

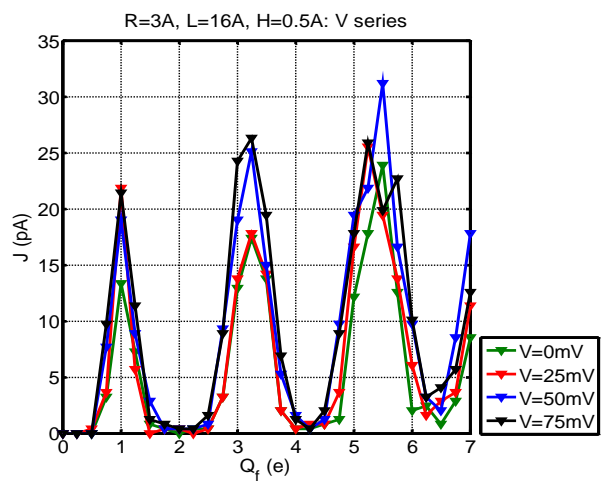

Figure A3. Calcium current $J$ vs fixed charge $Q_f$ showing how the band structure varies with the length of the charged cylindrical ring in the range $H$=0.5-8Å. The band structure is clearly not very sensitive to $H$.

Figure A4. Calcium current $J$ vs fixed charge $Q_f$ showing how the band structure depends on the membrane voltage $V$=0-75mV. It is evident that there is no significant variation with $V$.